%% file: main.tex
\documentclass[11pt]{article}

\usepackage[preprint]{acl}

\usepackage{listings}
\usepackage{times}
\usepackage{latexsym}
\usepackage[T1]{fontenc}
\usepackage[utf8]{inputenc}
\usepackage{microtype}
\usepackage{inconsolata}
\usepackage{graphicx}
\usepackage{enumitem}
\usepackage{booktabs}
\usepackage{longtable}
\usepackage{amssymb}
\usepackage{xspace}

\newcommand{\bench}{\textsc{BackendForge}\xspace}
\newcommand{\cmark}{\textcolor{green!60!black}{\ensuremath{\checkmark}}}
\newcommand{\xmark}{\textcolor{red!75!black}{\ensuremath{\times}}}

\title{\bench: Benchmarking Agentic End-to-End Code Generation with Backend Services}

\author{
\textbf{Yuzhe Guo\textsuperscript{1,2}\thanks{Equal contribution.}} \quad
\textbf{Mengzhou Wu\textsuperscript{1,2}\footnotemark[1]} \quad
\textbf{Yuan Cao\textsuperscript{1,2}} \quad
\textbf{Jialei Wei\textsuperscript{3}} \quad
\textbf{Dezhi Ran\textsuperscript{1,2}\thanks{Corresponding authors.}} \\
\textbf{Wei Yang\textsuperscript{3}} \quad
\textbf{Tao Xie\textsuperscript{1,2,3,4}\footnotemark[2]
} \\
\\
\textsuperscript{1}Key Lab of HCST (PKU), MOE; SCS, Peking University \\
\textsuperscript{2}Beijing Tongming Lake Information Technology Application Innovation Center
(TLAIC) \\
\textsuperscript{3}Fudan University Institute of Systems for Advanced Computing \\
\textsuperscript{4}Shanghai Institute of Systems for Open Computing \\
}

\begin{document}
\maketitle

\input{sections/abstract}

\input{sections/teaser}

\input{sections/introduction}

\input{sections/related_work}

\input{sections/benchmark_construction}

\input{sections/evaluation}

\input{sections/analysis}

\input{sections/conclusion}

\input{sections/limitation}
\bibliography{main}

\clearpage
\appendix
\input{sections/appendix}

\end{document}

%% file: sections/abstract.tex
\begin{abstract}
Large language models (LLMs) are increasingly used in agentic coding settings, where they can inspect files, execute commands, run tests, observe failures, and iteratively revise code.
This shift raises a central evaluation question: can an agentic LLM generate an end-to-end software artifact that is both deployable and behaviorally correct under execution?
Backend services provide a controlled but realistic substrate for this evaluation. Their APIs expose application-level executable semantics, and deployed behavior can be checked deterministically against an OpenAPI contract through black-box HTTP interactions.
We introduce \bench, a benchmark of 56 contract-defined backend generation tasks rewritten from real open-source applications.
Given a visible specification and an OpenAPI contract, an LLM must generate a Dockerized service that is built, deployed, and evaluated only through HTTP tests.
To strengthen evaluation without introducing hidden requirements, \bench uses a test agent and a code agent to co-evolve the test oracle and reference service, where the test agent proposes specification-grounded backend tests and the code agent repairs the reference implementation.
Although the best-performing model, GPT-5.5, succeeds on 55.4\% of tasks under the base oracle, it succeeds on only 28.6\% under the final oracle.
This gap suggests that current LLMs can implement many local API behaviors, but still struggle to produce complete backend services.
\end{abstract}

%% file: sections/teaser.tex
\begin{figure*}[t]
\centering
\includegraphics[width=0.98\textwidth]{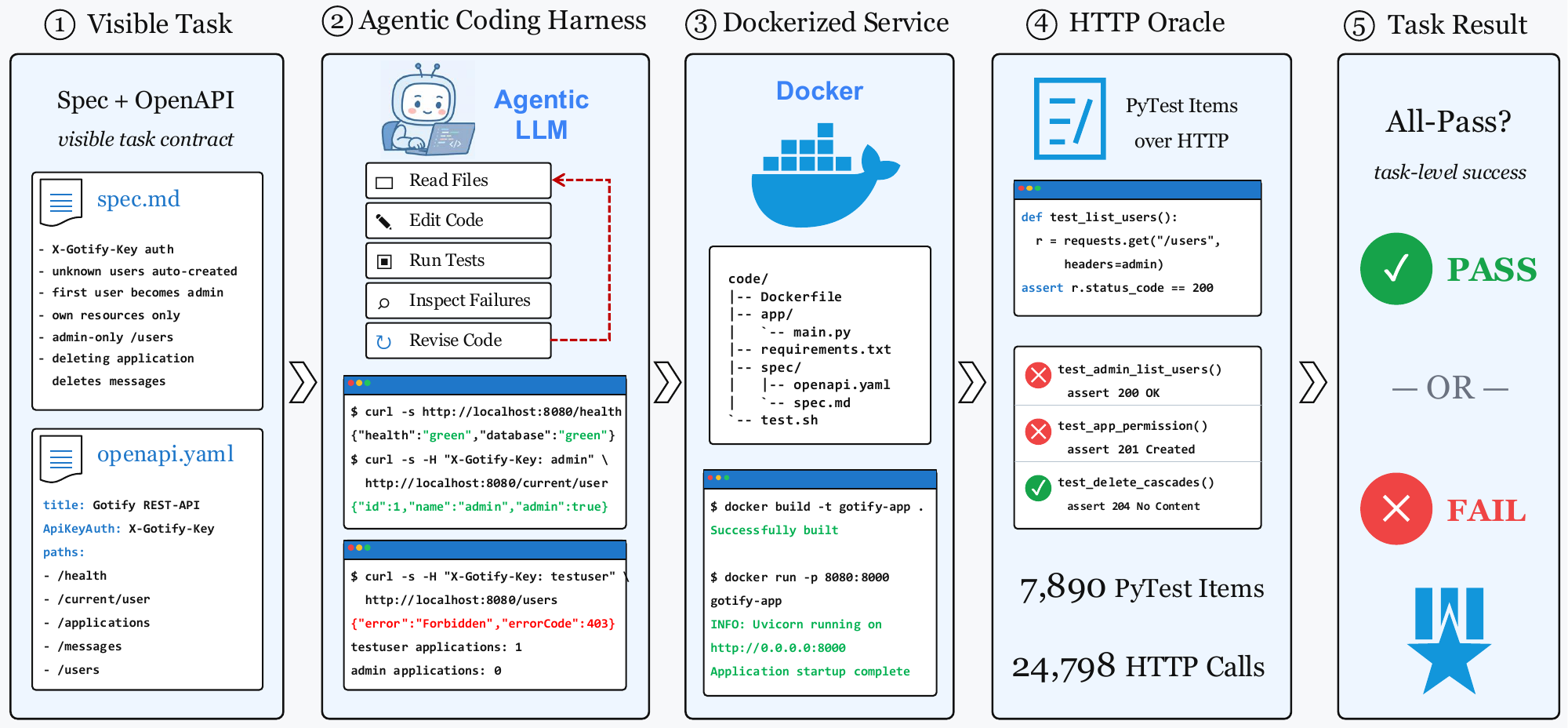}
\caption{\bench evaluation pipeline. The agent receives only the visible task contract, develops and self-tests a Dockerized backend service, and submits the service for black-box scoring. The evaluator deploys the submitted service and runs the hidden HTTP oracle. Task-level success requires all oracle items to pass.}
\label{fig:teaser}
\end{figure*}

%% file: sections/introduction.tex
\section{Introduction}

Large language models (LLMs) are rapidly changing the unit of software generation.
Early code generation systems primarily targeted localized assistance, such as completing a function, solving a programming problem, or producing a small code snippet~\citep{chen2021evaluating,austin2021program,hendrycks2021apps}.
Subsequent systems and benchmarks moved toward repository-level development, where models modify existing codebases to resolve bugs or implement requested changes~\citep{jimenez2024swebench}.
More recently, LLMs are used in agentic coding settings where they can inspect project files, execute commands, run tests, observe failures, and iteratively revise code within a workspace~\citep{yang2024sweagent,anthropic2026claudecode,opencode2026}.
As LLM coding workflows become more interactive and autonomous, user expectations are shifting from receiving isolated code fragments toward obtaining usable software artifacts~\citep{yao2023react,wu2024autogen,lu2021codexglue,hong2024metagpt}.
For such artifacts, usefulness depends not only on whether code compiles or local tests pass, but on whether the generated system can be built, deployed, and interacted with through its intended interfaces.

Evaluating end-to-end software generation requires balancing realism against deterministic evaluation~\citep{zhou2024webarena,xie2024osworld}.
A fully open-ended benchmark could ask models to infer requirements from underspecified user requests, design an interface, implement the system, and deploy it~\citep{qian2024chatdev,chen2024taskweaver}.
Such a setting would be realistic, but difficult to evaluate deterministically because failures could arise from requirement interpretation, interface design, implementation, or deployment~\citep{ezzini2021domainspecific,kapoor2024aiagents}.
In \bench, we instead isolate the contract-realization problem: given an explicit natural-language specification and an OpenAPI contract, can an agentic LLM construct a deployable backend service whose externally observable behavior satisfies that contract?

Existing benchmarks evaluate important components of agentic coding, but they do not directly evaluate whether an LLM can generate a deployable software service that behaves correctly after deployment.
Repository repair settings such as SWE-bench evaluate whether an LLM can modify an existing codebase to resolve a reported issue, requiring repository understanding and patch generation~\citep{jimenez2024swebench}.
Related agentic backend tasks adopt a similar formulation in backend settings, where models implement requested functionality inside existing repositories rather than generating deployable services from scratch~\citep{yang2026abcbench}.
Other work studies from-scratch application or backend generation~\citep{ran2025appforge,vero2025baxbench}, but typically evaluates single-turn standalone generation rather than deployable service construction under an agentic coding workflow.
Meanwhile, full-stack application benchmarks move closer to end-to-end software generation, but their evaluation often entangles service behavior with frontend rendering, browser automation, GUI interaction, and subjective or unstable interface-level signals~\citep{liu2026webcoderbench,tran2026vibecodebench}.

Backend services provide a controlled but realistic substrate for evaluating whether generated software behaves correctly after deployment~\citep{atlidakis2019restler,martinlopez2021restest}.
This choice is not meant to reduce end-to-end software generation to backend coding alone.
Rather, backend APIs expose much of an application's executable semantics, including interface contracts, persistent state, authentication, authorization, validation, cross-entity side effects, and workflows.
They also provide a machine-readable interface that LLMs in agentic workflows can directly invoke through structured requests and responses, reflecting a broader shift toward agent-native software interaction through APIs rather than graphical interfaces~\citep{schick2023toolformer,qin2024toolllm,patil2024gorilla}.
Unlike GUI-based evaluation~\citep{deng2023mind2web}, backend behavior can be grounded in explicit specifications, OpenAPI contracts~\citep{openapi2024oas}, and deterministic HTTP interactions~\citep{viglianisi2020resttestgen}.
Backend services therefore provide a useful testbed for isolating application-level behavior while still requiring generated code to run as a deployed system.

We introduce \bench, a benchmark for evaluating deployable backend service generation under agentic coding.
Each task provides a visible natural-language specification together with an OpenAPI contract, and an LLM must generate a backend service that can be built, deployed, and evaluated through black-box HTTP interactions.
\bench contains 56 tasks rewritten from real open-source applications across domains such as e-commerce, content management, project collaboration, notification systems, finance tools, identity management, feature flags, and developer tooling.
To reduce direct recoverability from upstream repositories, we rewrite the API contracts, authentication schemes, implementation stacks, and test suites while preserving realistic service semantics.

A central challenge in this setting is constructing an oracle that is strong enough to expose service-level defects without introducing hidden requirements~\citep{barr2015oracle,alonso2023agora}.
Backend services contain many corner cases involving state consistency, authorization isolation, validation, filtering, side effects, and multi-step workflows.
Seed test suites rarely cover these behaviors exhaustively, but automatically generated tests can over-specify the task by adding requirements not entailed by the visible specification~\citep{fraser2015unit}.
To address this tension, \bench uses multi-agent oracle co-evolution during benchmark construction.
Test generation, specification validation, and reference repair co-evolve under the visible specification and OpenAPI contract.
A candidate test is admitted only if it exposes a specification-grounded defect in the reference implementation and the repaired reference passes full regression.
This process strengthens oracle coverage of backend corner cases while preserving the task contract given to the model.

Our evaluation shows that current LLMs can implement many local API behaviors, but still struggle to produce complete deployable backend services.
We report the success rate, defined as the fraction of tasks for which the generated candidate service passes the final oracle.
Across 56 tasks, the strongest model succeeds on only 16 tasks (28.6\%).
The failures are not dominated by syntax errors or missing endpoints.
Instead, models most often fail on service-level execution semantics, including state consistency, authorization isolation, validation, side-effect propagation, and multi-step workflows.
These results suggest that deployable backend service generation remains a challenging benchmark for LLMs in agentic coding workflows.

Our contributions are as follows:
\begin{itemize}[leftmargin=*]
    \item We develop a scalable methodology for constructing specification-grounded backend service oracles, using multi-agent test generation, specification validation, and reference repair to harden HTTP test suites without introducing hidden requirements.

    \item We introduce \bench, a 56-task benchmark for end-to-end backend service generation from explicit natural-language specifications and OpenAPI contracts, paired with a deterministic black-box HTTP verifier for deployed service behavior.

    \item We present a systematic empirical evaluation of current LLMs on \bench, characterizing task-level success and recurring semantic failure modes in deployable backend generation.
\end{itemize}

%% file: sections/related_work.tex
\section{Related Work}

\begin{table}[t]
\centering
\scriptsize
\setlength{\tabcolsep}{3pt}
\begin{tabular}{@{}lccc@{}}
\toprule
Benchmark & Rigorous Eval. & Agentic & End-to-End \\
\midrule
HumanEval~\citep{chen2021evaluating} & \cmark & \xmark & \xmark \\
SWE-bench~\citep{jimenez2024swebench} & \cmark & \cmark & \xmark \\
WebCoderBench~\citep{liu2026webcoderbench} & \xmark & \cmark & \cmark \\
Vibe Code Bench~\citep{tran2026vibecodebench} & \xmark & \cmark & \cmark \\
AppForge~\citep{ran2025appforge} & \xmark & \xmark & \cmark \\
BaxBench~\citep{vero2025baxbench} & \cmark & \xmark & \cmark \\
ABC-Bench~\citep{yang2026abcbench} & \cmark & \cmark & \xmark \\
\bench (ours) & \cmark & \cmark & \cmark \\
\bottomrule
\end{tabular}
\caption{Comparison with representative code, application, and backend benchmarks. Rigorous Eval. denotes deterministic scoring. Agentic denotes interaction with a development environment. End-to-End denotes generation of a runnable application or service from a specification.}
\label{tab:benchmark-comparison}
\end{table}

\bench differs from prior benchmarks by combining three properties that are usually studied separately: reproducible evaluation, agentic development, and end-to-end runnable output.
Table~\ref{tab:benchmark-comparison} compares representative benchmarks along these three axes.
Unlike benchmarks that focus on isolated functions, repository-level patches, or application generation without deterministic scoring, \bench evaluates whether an agent-generated backend service satisfies a visible API contract after deployment under a co-evolved HTTP oracle.

\paragraph{Code generation and repository repair.}
Code generation benchmarks evaluate functions, classes, or programming-problem solutions under unit tests or execution checks~\citep{chen2021evaluating,austin2021program,hendrycks2021apps,du2023classeval}, with later work strengthening the oracle through augmented or generated tests~\citep{liu2023evalplus,chen2022codet}.
Repository repair benchmarks such as SWE-bench evaluate patch generation and agent-computer interaction within existing codebases~\citep{jimenez2024swebench,yang2024sweagent}.

\begin{figure*}[htbp]
\centering
\includegraphics[width=\textwidth]{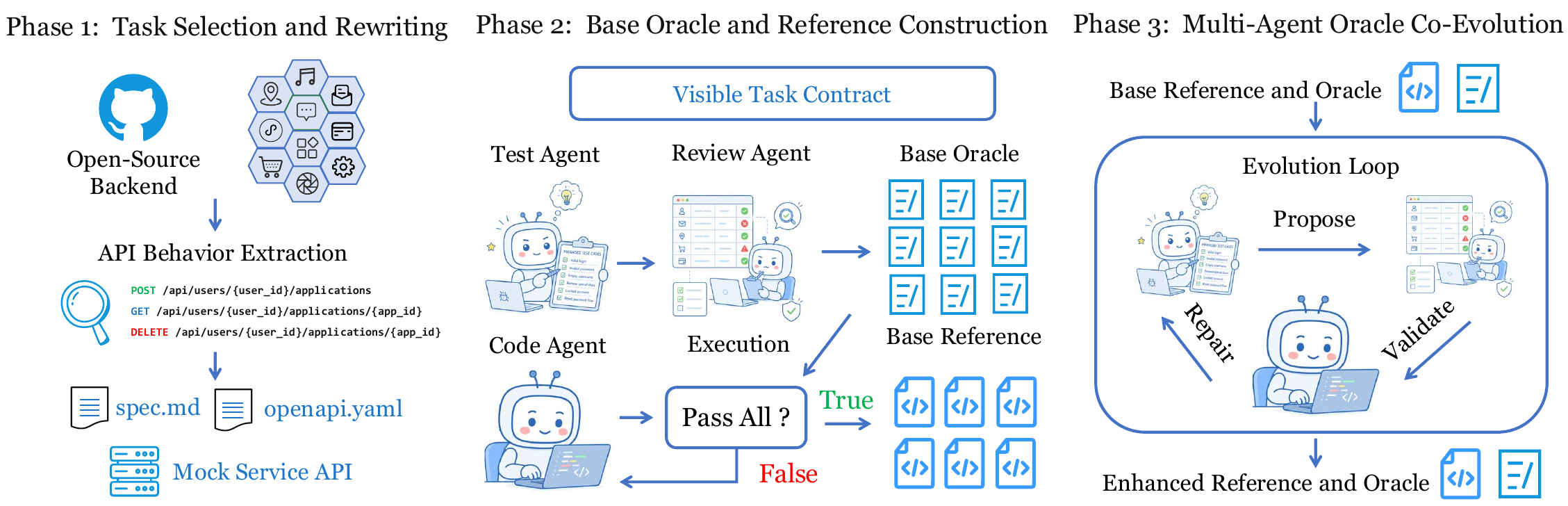}
\caption{\bench construction pipeline. Source backend applications are rewritten into visible task contracts, which serve as the source of truth for base oracle construction, reference construction, and repair-guided oracle co-evolution. Candidate oracle items are admitted only when they are grounded in the visible contract and expose a defect in the old reference.}
\label{fig:construction-pipeline}
\end{figure*}

\paragraph{Application and service generation.}
WebCoderBench, Vibe Code Bench, and AppForge evaluate web or application generation from user-facing requirements~\citep{liu2026webcoderbench,tran2026vibecodebench,ran2025appforge}.
Their evaluation can mix backend logic with frontend rendering, browser workflows, visual correctness, and platform-specific constraints.

\paragraph{Backend generation and agentic backend coding.}
BaxBench evaluates standalone backend application generation from scenario specifications across frameworks~\citep{vero2025baxbench}, and ABC-Bench evaluates agentic backend coding inside existing repositories with environment setup and external API tests~\citep{yang2026abcbench}.

%% file: sections/benchmark_construction.tex
\section{\bench}
\label{sec:benchmark-construction}
\bench is built around three invariants that make end-to-end backend generation measurable. First, every task has a visible contract consisting of a natural-language specification and an OpenAPI file. Second, every scored artifact is a deployed service observed only through HTTP requests. Third, oracle strengthening is allowed only when a candidate test is grounded in the visible contract.
Figure~\ref{fig:construction-pipeline} summarizes this construction pipeline. Rather than trusting a single LLM-generated artifact, \bench uses LLM agents only in roles where their outputs can be reviewed, executed, and repaired under deterministic feedback.

\bench contains 56 tasks with substantial API surface area and test coverage. Across these tasks, the OpenAPI contracts define 2,345 API operations, where an operation is an HTTP method and path pair such as \texttt{GET /users/\{id\}} or \texttt{POST /projects}. The final oracle contains 7,890 pytest items, consisting of 7,250 base oracle items and 640 admitted co-evolved pytest items. A pytest item is one collected pytest test unit after parameterization, and each item issues one or more HTTP requests against the deployed service. In total, the oracle executes 24,798 HTTP request invocations, so \bench measures stateful service behavior rather than isolated function outputs.

\subsection{Task Selection and Controlled Rewriting}

\bench starts from real open-source backend applications, but each source is rewritten into a controlled benchmark task. We select applications that exercise service-level backend behavior, especially persistence, access control, validation, entity relationships, and multi-step workflows. The resulting task set spans commerce, collaboration, productivity, identity, analytics, and infrastructure-style domains. For more details, see Appendix~\ref{app:task-list}.
This selection keeps the benchmark close to realistic backend engineering while avoiding tasks that reduce to a single algorithmic routine or a static response template.

Controlled rewriting serves two purposes by preserving realistic service semantics while preventing upstream code from being a drop-in solution. 
For every task, we rewrite the API surface, authentication model, behavior rules, and evaluation items, then express the rewritten behavior through a visible natural-language specification and an OpenAPI contract.
The original application code therefore cannot simply be copied to pass the benchmark. 
At the same time, the task remains solvable from the released materials because the required behavior is documented in the visible contract rather than hidden in the source repository.

\subsection{Visible Task Contract}

Each \bench task exposes a visible contract that is the single source of truth for construction and evaluation. The visible materials consist of a product specification (\texttt{spec.md}) and an OpenAPI contract (\texttt{openapi.yaml}). The specification describes behavior that is difficult to capture in schemas alone, such as roles, ownership rules, state transitions, and workflow side effects. The OpenAPI contract defines the endpoint surface, request and response schemas, status codes, and authentication mechanism. During construction, every generated test, review decision, and reference repair is checked against these two artifacts rather than against private implementation assumptions.

The visible contract also defines the execution boundary of a task. A submitted solution must provide a Python backend service and a Dockerfile that builds and runs the service in isolation. We restrict the implementation language to Python to control the experimental setting, but leave framework, storage, and internal design choices open. The benchmark does not inspect source code during scoring. It evaluates only HTTP behavior at the service boundary, allowing different internal implementations to be compared under the same behavioral contract.

Tasks that require external integrations use deterministic mock services rather than real third-party APIs. For example, calendar booking, notification delivery, and payment workflows may require the generated service to call a harness-provided HTTP service under \texttt{mock\_services/}. These mocks are part of the benchmark environment, not part of the candidate output. The visible \texttt{spec.md} describes the mock service URL and request/response contract, so the service must integrate with it as it would with an external API. In the current task set, five tasks include such mock services for calendar, delivery, or payment behavior. This design keeps integration-heavy backend semantics in scope while avoiding nondeterministic online dependencies.

\subsection{Base Oracle and Reference Construction}

The base oracle is constructed before the reference implementation so that the first regression target is not fitted to a particular service. 
During task rewriting, a test agent receives \texttt{spec.md} and \texttt{openapi.yaml}, then drafts black-box pytest items for documented endpoint behavior and multi-step workflows. 
A review agent checks every candidate item against the same two visible materials and rejects items that overinterpret the task, guess undocumented behavior, or rely on speculative business rules. 
The retained items become the fixed base oracle. Across the 56 tasks, this process produces 7,250 base oracle items that establish broad coverage over endpoint behavior and common workflows.

Only after the base oracle is fixed do we build the reference implementation. A code agent receives \texttt{spec.md} and \texttt{openapi.yaml} as the implementation target, while the fixed base oracle acts as a regression suite for the current reference. When the reference fails, the construction harness reports the failing pytest item and a concise error summary, allowing the code agent to repair local implementation defects. 

\begin{table*}[htbp]
\centering
\small
\resizebox{\textwidth}{!}{
\begin{tabular}{lcccccccc}
\toprule
LLM (Thinking Effort) & Size & Base SR & +R1 SR & +R2 SR & Final SR & $\Delta$ SR & Cost/Task & Output Tokens/Task \\
\midrule
\multicolumn{9}{l}{\emph{Proprietary LLMs}} \\
\midrule
GPT-5.5 (xhigh) & N/A & 55.36\% & \textbf{42.86\%} & \textbf{37.50\%} & \textbf{28.57\%} & \textcolor{red}{-48.39\%} & \$3.39 & 51.06K \\
Claude Opus 4.7 (max) & N/A & \textbf{58.93\%} & 37.50\% & 28.57\% & 17.86\% & \textcolor{red}{-69.70\%} & \$5.55 & 75.38K \\
Gemini 3.5 Flash & N/A & 21.43\% & 10.71\% & 8.93\% & 5.36\% & \textcolor{red}{-75.00\%} & \$2.15 & 61.30K \\
Claude Sonnet 4.6 (max) & N/A & 39.29\% & 8.93\% & 5.36\% & 3.57\% & \textcolor{red}{-90.91\%} & \$2.76 & 65.17K \\
Qwen 3.6 Max (high) & N/A & 23.21\% & 5.36\% & 1.79\% & 1.79\% & \textcolor{red}{-92.31\%} & \$0.73 & 42.04K \\
Qwen 3.6 Plus (high) & N/A & 23.21\% & 1.79\% & 0.00\% & 0.00\% & \textcolor{red}{-100.00\%} & \$0.33 & 43.91K \\
\midrule
\multicolumn{9}{l}{\emph{Open-Source LLMs}} \\
\midrule
DeepSeek V4 Pro (max) & 1.6T-A49B & 32.14\% & 12.50\% & 7.14\% & 3.57\% & \textcolor{red}{-88.89\%} & \$0.09 & 50.85K \\
GLM 5.1 & 754B-A40B & 26.79\% & 8.93\% & 3.57\% & 3.57\% & \textcolor{red}{-86.67\%} & \$1.31 & 54.82K \\
Kimi K2.6 & 1T-A32B & 39.29\% & 12.50\% & 3.57\% & 1.79\% & \textcolor{red}{-95.45\%} & \$0.77 & 52.54K \\
DeepSeek V4 Pro (high) & 1.6T-A49B & 25.00\% & 8.93\% & 3.57\% & 1.79\% & \textcolor{red}{-92.86\%} & \$0.07 & 42.34K \\
DeepSeek V4 Flash (high) & 284B-A13B & 21.43\% & 3.57\% & 0.00\% & 0.00\% & \textcolor{red}{-100.00\%} & \$0.03 & 40.24K \\
DeepSeek V4 Flash (max) & 284B-A13B & 19.64\% & 8.93\% & 1.79\% & 0.00\% & \textcolor{red}{-100.00\%} & \$0.04 & 53.25K \\
Hunyuan Hy3 (high) & 295B-A21B & 12.50\% & 3.57\% & 0.00\% & 0.00\% & \textcolor{red}{-100.00\%} & \$0.27 & 48.79K \\
MiniMax M2.7 & 229B-A10B & 5.36\% & 1.79\% & 0.00\% & 0.00\% & \textcolor{red}{-100.00\%} & \$0.27 & 48.50K \\
\bottomrule
\end{tabular}
}
\caption{Main \bench results. SR denotes task-level all-pass success rate over the same 56-task denominator. +R1 and +R2 cumulatively add accepted co-evolved items to the base oracle; the final oracle adds all accepted co-evolved items. $\Delta$ is the signed relative change from Base SR to Final SR, computed as $(\mathrm{Final\ SR} - \mathrm{Base\ SR}) / \mathrm{Base\ SR}$; red negative values indicate relative decreases. Closed LLMs and LLMs with undisclosed sizes use N/A in the Size column. A denotes active parameters for MoE LLMs.}
\label{tab:main-results}
\end{table*}

\subsection{Multi-Agent Oracle Co-Evolution}

The base oracle is broad, but it is not treated as complete. Backend correctness often depends on authorization boundaries, state isolation, validation, reference integrity, and workflow side effects that are easy to miss in the initial base oracle construction. At the same time, automatically adding harder tests is unsafe because a plausible backend behavior is not necessarily a specified behavior. For example, one API may return \texttt{404} when deleting a missing resource, while another treats deletion as idempotent and returns \texttt{204}. If the visible contract does not choose one behavior, a pytest item that enforces either choice introduces a hidden requirement. The co-evolution loop addresses this tension by searching for useful missing tests while using review and repair to prevent unsupported requirements from entering the oracle.

The co-evolution loop uses the current reference as a probe for missing oracle coverage. Given the visible contract, the current reference implementation, and the already accepted items, the test agent proposes a candidate black-box pytest item and executes it against the old reference. If the candidate passes, it is discarded in this loop because it does not reveal a reference defect. If the candidate fails, the candidate and its reference-execution trace are passed to the review agent. This failure-first filter makes the loop scalable. Many candidates can be generated, but only candidates that expose concrete reference behavior receive further attention.

The review agent turns a failing candidate into either a rejected assumption or an eligible reference defect. It checks whether the candidate has traceable support in \texttt{spec.md} or \texttt{openapi.yaml}, and rejects items that depend on undocumented behavior, ambiguous status-code choices, or harness mistakes. The remaining failures are treated as genuine reference defects. This review step is necessary because failure on the old reference alone is not evidence that a test belongs in the benchmark.

Confirmed reference defects are then passed to the code agent for constrained repair.
The code agent may modify only the task reference implementation, not the specification, OpenAPI contract, or reviewed candidate tests.
Once a candidate item has passed review, it is treated as a valid specification-grounded requirement, and the reference is repaired until it satisfies this item together with the base oracle and all previously accepted evolved items.
After each repair attempt, the harness runs the full regression suite against the patched reference.
On each failed attempt, the harness reports the failing pytest item and a concise error summary to guide the next repair.
The item is admitted only after the repaired reference passes this full regression suite, ensuring that oracle growth preserves all existing accepted behavior.

Accepted items are accumulated to form the co-evolved part of the final oracle. The final \bench artifact contains the rewritten visible contracts, deterministic mock services when needed, internal reference implementations used during construction, and the final oracle. The final oracle is the union of the 7,250 base oracle items and 640 admitted co-evolved items. This construction uses LLM agents where their outputs can be checked locally and repaired under regression, while the benchmark boundary remains deterministic. Only reviewed, contract-grounded, regression-preserving items become part of the scored oracle.

%% file: sections/evaluation.tex
\section{Evaluation}
\label{sec:evaluation}

\subsection{Experimental Setup}

We select frontier agentic LLMs from both proprietary and open-source families, and evaluate all of them using the same mini-SWE-style agentic coding harness based on the agent-computer-interface design principles studied by SWE-agent~\citep{yang2024sweagent}.
The proprietary models are GPT-5.5~\citep{openai2026gpt55}, Claude Opus 4.7~\citep{anthropic2026claudeopus47}, Gemini 3.5 Flash~\citep{google2026gemini35flash}, Claude Sonnet 4.6~\citep{anthropic2026claudesonnet46}, Qwen 3.6 Max~\citep{qwen2026qwen36max}, and Qwen 3.6 Plus~\citep{qwen2026qwen36plus}.
The open-source models are DeepSeek V4 Pro~\citep{deepseek2026v4techreport}, GLM 5.1~\citep{zai2026glm5techreport}, Kimi K2.6~\citep{moonshot2026kimik26}, DeepSeek V4 Flash~\citep{deepseek2026v4techreport}, Hunyuan Hy3~\citep{tencent2026hy3}, and MiniMax M2.7~\citep{minimax2026m27hf}.


We report task-level success rate (SR) as the primary metric.
For each oracle, a task is counted as successful only if the generated service passes all required pytest items for that task.


\subsection{Main Results}

\textbf{Current LLMs generate many locally correct backend behaviors, but they rarely construct fully correct services.} Table~\ref{tab:main-results} reports task-level success rates over the same 56 tasks. Under the base oracle, the strongest LLMs solve a substantial fraction of tasks: Claude Opus 4.7 reaches 58.9\%, and GPT-5.5 reaches 55.4\%. Under the final oracle, however, success drops sharply. GPT-5.5 solves 16 of 56 tasks (28.6\%), Claude Opus 4.7 solves 10 (17.9\%), and every other configuration solves at most 3. This gap shows that endpoint-level progress is real but incomplete. Generated services often handle many routes and workflows while still failing one required invariant, validation rule, or cross-entity side effect.


\textbf{Top proprietary models provide the strongest final-oracle performance at substantially higher generation cost.}
GPT-5.5 and Claude Opus 4.7 solve 16/56 and 10/56 tasks under the final oracle, while every other configuration solves at most 3/56.
This performance advantage comes with higher average cost per task. GPT-5.5 costs \$3.39 and Claude Opus 4.7 costs \$5.55 in our runs, compared with sub-dollar costs for several open-source configurations.

%% file: sections/analysis.tex
\section{Analysis}
\label{sec:analysis}


\subsection{A small hardened oracle has a large effect}

\begin{figure}[t]
\centering
\includegraphics[width=\columnwidth]{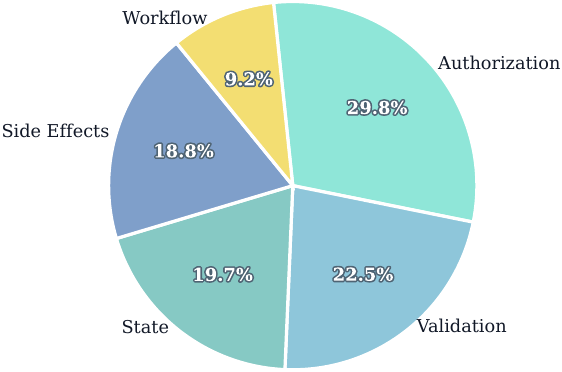}
\caption{Semantic distribution of accepted co-evolved pytest items in the final oracle. Each accepted co-evolved pytest item is assigned to one primary category using its recorded bug-pattern metadata.}
\label{fig:pytest-item-categories}
\end{figure}

\begin{table}[t]
\centering
\small
\setlength{\tabcolsep}{2.2pt}
\begin{tabular}{lrrrrr}
\toprule
Category & Base & Base \% & Evol. & Evol. \% & $\Delta$ pp \\
\midrule
API surface & 2,636 & 36.4 & 0 & 0.0 & $-36.4$ \\
Authorization & 1,041 & 14.4 & 191 & 29.8 & $+15.5$ \\
Validation & 2,248 & 31.0 & 144 & 22.5 & $-8.5$ \\
State & 177 & 2.4 & 126 & 19.7 & $+17.2$ \\
Side effects & 709 & 9.8 & 120 & 18.8 & $+9.0$ \\
Workflow & 439 & 6.1 & 59 & 9.2 & $+3.2$ \\
\midrule
Total & 7,250 & 100.0 & 640 & 100.0 & -- \\
\bottomrule
\end{tabular}
\caption{Distributional contrast between base oracle and accepted co-evolved pytest items. $\Delta$ is Evolved \% minus Base \%.}
\label{tab:base-evolve-distribution}
\end{table}

\begin{figure}[t]
\centering
\includegraphics[width=\columnwidth]{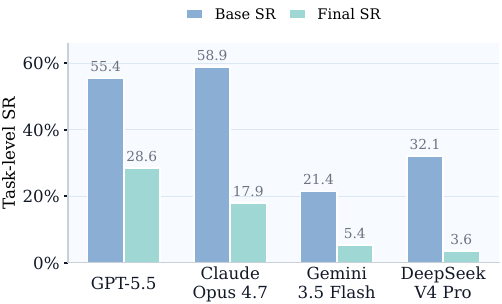}
\caption{Success-rate drop from the base oracle to the final oracle for representative LLMs.}
\label{fig:oracle-attrition}
\end{figure}

\begin{figure}[t]
\centering
\includegraphics[width=\columnwidth]{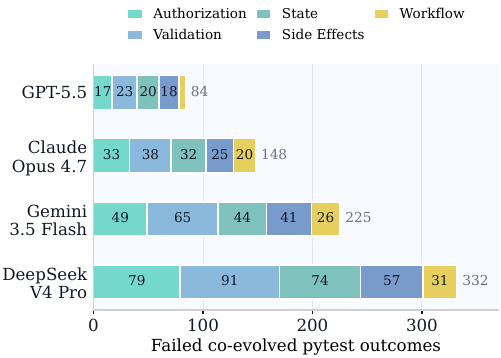}
\caption{Distribution of error types across models.}
\label{fig:error-type-distribution}
\end{figure}

\textbf{The hardened oracle adds relatively few pytest items, but these items have a large effect on task-level success.}
Table~\ref{tab:base-evolve-distribution} shows that the final oracle contains 640 accepted co-evolved pytest items in addition to the 7,250 base oracle items. This is a modest 8.8\% increase in item count, yet Table~\ref{tab:main-results} shows a large decrease in all-pass success once those items are included. The added items are not random extra coverage. Figure~\ref{fig:pytest-item-categories} shows that they concentrate on authorization, validation, and stateful workflow behavior. 


\textbf{The retained co-evolved items reveal failures that the base oracle misses.}
Figure~\ref{fig:oracle-attrition} shows monotonic attrition from the base oracle to the final oracle.
Claude Opus 4.7 drops from 58.9\% to 17.9\%, while GPT-5.5 drops from 55.4\% to 28.6\%.
These drops indicate that the accepted co-evolved items capture specification-grounded behaviors beyond broad endpoint coverage.
Complete backend success therefore requires satisfying many small contract semantics simultaneously.

\textbf{The remaining failures cover multiple backend contract categories.}
Figure~\ref{fig:error-type-distribution} shows that validation is the largest category across models, ranging from 23 failed outcomes for GPT-5.5 to 91 for DeepSeek V4 Pro.
Authorization and state failures are also substantial, with DeepSeek V4 Pro failing 79 authorization outcomes and 74 state outcomes.
Thus, final-oracle attrition reflects broad difficulty with ordinary service contracts rather than one narrow failure mode.


\subsection{Review Agent Prevents Hidden Requirements}

\textbf{The review agent is essential to making the hardened oracle a stable benchmark signal rather than merely a harder test suite.} It reviews candidate pytest items only after the test agent has shown that they fail on the current reference implementation. The review agent then checks the failing candidates against the visible specification and OpenAPI contract, rejecting items that require behavior not uniquely supported by those materials. In final pool construction, it filtered 113 candidate pytest items. The most common reasons were task overinterpretation, unsupported assertions, and written-contract gaps.


\subsection{Hidden Self-Verification Helps but Does Not Solve the Benchmark}

\begin{figure}[t]
\centering
\includegraphics[width=\columnwidth]{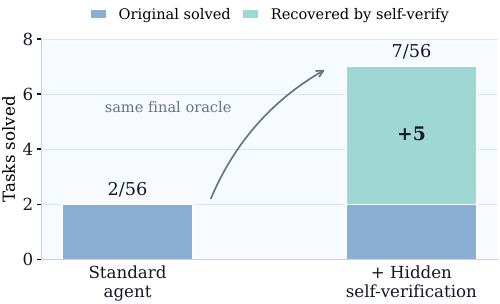}
\caption{Hidden self-verification improves DeepSeek V4 Pro on the final oracle.}
\label{fig:self-verify-gain}
\end{figure}

\textbf{Sparse hidden feedback improves final-oracle success.}
We evaluate a stronger protocol in which the agent can call \texttt{backendforge\_self\_verify} during development.
The tool builds the current candidate service and runs the final oracle, but returns only \texttt{SELF\_VERIFY\_PASS} or \texttt{SELF\_VERIFY\_FAIL} with a coarse failure phase such as build, startup, or tests.
It does not expose pytest node IDs, assertions, hidden requests, test source code, the reference implementation, or construction-time traces.
Under this protocol, DeepSeek V4 Pro solves 7 of 56 tasks under the final oracle, compared with 2 of 56 in the main protocol.
As shown in Figure~\ref{fig:self-verify-gain}, the original two successes are retained, and self-verification recovers five additional tasks.

\begin{figure}[t]
\centering
\setlength{\fboxsep}{4pt}
\fcolorbox{black!15}{black!2}{%
\begin{minipage}{0.92\columnwidth}
\scriptsize\ttfamily
[verifier call 9]\\
SELF\_VERIFY\_FAIL\\
Phase: tests\\[2pt]
[LLM thinking]\\
I've tried many fixes but I'm still failing...\\
OK, I've done everything I can think of.\\
I'm going to stop here and submit.\\[2pt]
[tool call]\\
echo COMPLETE\_TASK\_AND\_SUBMIT\_FINAL\_OUTPUT
\end{minipage}}
\caption{Representative DeepSeek V4 Pro self-verification trace. In the \texttt{maybefinance} task, the agent receives a ninth test-phase \texttt{FAIL} and submits without a \texttt{PASS} signal, despite remaining verifier budget.}
\label{fig:self-verify-abandonment}
\end{figure}

\textbf{Agents may abandon exploration even when hidden feedback remains available.}
Figure~\ref{fig:self-verify-abandonment} shows a representative failure trajectory.
After nine failed verifier calls, DeepSeek V4 Pro recognizes that the service is still failing, exhausts its own debugging hypotheses, and submits without receiving a \texttt{SELF\_VERIFY\_PASS} signal.
Thus, sparse hidden feedback changes the development protocol and can recover some failures, but it does not collapse the benchmark. 49 of 56 tasks still fail for this model even with self-verification.

%% file: sections/conclusion.tex
\section{Conclusion} As LLM coding workflows become increasingly agentic, evaluation must move beyond isolated code generation toward deployable end-to-end software artifacts. 
\bench approaches this problem through backend contract realization, where a visible specification and OpenAPI contract define a deployable service evaluated through deterministic black-box HTTP interactions.
Using a multi-agent oracle co-evolution framework, we strengthen backend evaluation without introducing hidden requirements. 
Our results show that current LLMs can implement many local API behaviors, but still struggle with core backend semantics such as validation, authorization, state consistency, side effects, and workflows. 
Additional experiments indicate that self-verification can partially mitigate these failures, but most tasks remain unsolved, suggesting that deployable backend generation remains a challenging problem for agentic LLMs.

%% file: sections/limitation.tex
\section*{Limitations and Potential Risks}

\bench is scoped to Python backend services evaluated through deterministic black-box HTTP tests. 
This scope makes the benchmark controllable and reproducible, but it does not cover full-stack user interfaces, other implementation stacks, performance engineering, deployment security, or production hardening. 

The final oracle is strengthened through repair-guided co-evolution, but it is still bounded by the visible task contracts and by the admitted test items. 
Although accepted co-evolved items are reviewed for contract grounding and regression preservation, the oracle cannot exhaustively cover all possible backend behaviors. 
A model that passes \bench should therefore not be assumed to produce production-ready services without additional review, security analysis, and deployment-specific testing.

The benchmark also carries potential dual-use risks. 
Improving agentic backend generation may help automate useful software development, but the same capability could be misused to generate backend services for spam, abuse infrastructure, deceptive applications, or poorly secured data-handling systems. 
For this reason, \bench evaluates functional correctness under visible contracts rather than endorsing unrestricted autonomous deployment. 
Systems built with such agents should include human oversight, security checks, access-control review, and safeguards before being connected to real users, sensitive data, or external services.

%% file: sections/appendix.tex
\section{Agent Prompt Templates}
\label{app:agent-prompts}

This appendix reports the prompt templates used by the construction agents in
\bench. The templates are lightly normalized for presentation: task-specific
content, previous-round summaries, and failure reports are shown as placeholders,
while operational details unrelated to the paper's construction logic are
omitted. The three prompts reflect the roles in Section~\ref{sec:benchmark-construction}:
the test agent proposes black-box pytest items, the code agent implements or
repairs the reference service, and the review agent decides whether a candidate
test is grounded in the visible contract.

\subsection{Test Agent}

The test agent receives the visible task contract and construction history, then
proposes candidate pytest items. The prompt emphasizes black-box HTTP behavior
and requires every assertion to be traceable to \texttt{spec.md} or
\texttt{openapi.yaml}.

\begin{lstlisting}[caption={Test agent prompt template.},label={lst:test-agent-prompt}]
You are the BackendForge test agent.

Goal:
Generate adversarial black-box HTTP pytest tests for a backend service
described by spec.md and openapi.yaml.

Authority:
- Use only the visible specification and OpenAPI contract.
- Do not invent hidden requirements.
- Every assertion must be traceable to spec.md or openapi.yaml.

Focus on backend contract failures:
1. field naming or response-shape mismatch
2. missing nested relation data
3. declared filters, sorting, or pagination not implemented
4. incorrect HTTP status codes
5. empty or fake business logic
6. authorization or data-isolation failure
7. cascade, side-effect, or preservation failure
8. uniqueness, ordering, idempotency, or state-machine failure

Inputs:
- Task name: {{ task_name }}
- spec.md: {{ spec_md }}
- openapi.yaml: {{ openapi_yaml }}
- Existing base tests: {{ existing_tests_summary }}
- Previously accepted co-evolved tests: {{ accepted_tests_summary }}
- Previously rejected candidate tests: {{ rejected_tests_summary }}
- Reference/testcase repair decisions: {{ reference_decision_summary }}
- Round guidance: {{ round_guidance }}

Rules:
- Use HTTP requests against BASE_URL.
- Do not inspect source code, database files, logs, or internal state.
- Avoid arbitrary timing assumptions and nondeterministic ordering.
- Prefer tests likely to catch realistic LLM-generated backend bugs.
- Do not repeat a previously rejected unsupported assertion.
- If prior accepted tests passed, search for deeper workflows, negative cases,
  ownership matrices, state transitions, and side-effect invariants.

Return only JSON:
{
  "candidates": [
    {
      "test_name": "...",
      "level": "L2 or L3",
      "bug_patterns": ["authorization_isolation"],
      "spec_traces": ["exact spec quote or section"],
      "openapi_traces": ["GET /api/path response schema"],
      "rationale": "...",
      "expected_reference_behavior": "...",
      "expected_generated_failures": ["run_name: reason"],
      "test_code": "standalone pytest code using requests and BASE_URL"
    }
  ]
}
\end{lstlisting}

\subsection{Code Agent}

The code agent is used during benchmark construction to build and repair the
reference service. Its prompt separates implementation from oracle definition:
the agent may repair only the reference implementation, while the visible
contract and accepted tests remain fixed.

\begin{lstlisting}[caption={Code agent prompt template for reference implementation and repair.},label={lst:code-agent-prompt}]
You are the BackendForge code agent.

Goal:
Implement or repair the task reference backend so that it satisfies the visible
contract in spec.md and openapi.yaml.

Authority:
- The visible specification and OpenAPI contract define the required behavior.
- Test failures are debugging evidence, not permission to hardcode a testcase.
- If a requested behavior is not supported by the visible contract, mark it for
  review instead of patching around it.

Allowed write scope:
- tasks/{{ task_name }}/reference/**

Forbidden changes:
- Do not modify spec.md.
- Do not modify openapi.yaml.
- Do not modify existing tests, candidate tests, or co-evolution artifacts.
- Do not depend on generated model outputs.

Implementation requirements:
- Build a complete backend service that listens on port 8000.
- Initialize required storage on startup, but do not insert seed data.
- Match documented endpoints, parameters, status codes, and response schemas.
- Implement general service behavior, not special cases for candidate values,
  test function names, UUIDs, emails, or resource names.
- Preserve authorization, ownership, validation, uniqueness, workflow, and
  side-effect semantics required by the visible contract.

Repair inputs:
- Task name: {{ task_name }}
- spec.md: {{ spec_md }}
- openapi.yaml: {{ openapi_yaml }}
- Repair batch: {{ repair_batch_json }}

After repair, all of the following must pass:
1. original base oracle tests
2. previously accepted co-evolved tests
3. current candidate tests in the repair batch
4. fresh-container regression checks

Output:
- files changed
- claims or failures addressed
- behavior implemented
- regression commands run
- any candidate that should be reclassified as SPEC_GAP or TESTCASE_BUG
\end{lstlisting}

\subsection{Review Agent}

The review agent protects the benchmark from hidden requirements. It reviews the
behavioral claim of each candidate test rather than incidental setup code, and
keeps only tests whose central assertion is directly supported by the visible
contract or is an unavoidable consequence of it.

\begin{lstlisting}[caption={Review agent prompt template.},label={lst:review-agent-prompt}]
You are the BackendForge review agent.

Goal:
Decide whether each candidate pytest item is grounded in the visible backend
contract and can be admitted to the benchmark oracle.

Authority:
- Use only spec.md and openapi.yaml.
- Do not use the reference implementation as a source of requirements.
- Do not use generated implementation behavior, product knowledge, framework
  conventions, or subjective engineering preferences as requirements.

Review standard:
- Review the behavioral purpose of the candidate, not incidental setup code.
- KEEP only if the asserted requirement is directly stated by spec.md,
  directly stated by openapi.yaml, or an unavoidable consequence of an explicit
  contract.
- FILTER if the candidate relies on plausible but undocumented behavior, an
  ambiguous status code, unspecified ordering, default pagination behavior,
  unstated cascade effects, implementation details, or a stronger rule than the
  contract supports.
- Do not filter merely because a test is negative or strict. If the rejection
  rule is explicit, checking that a rejected request creates no successful
  mutation is allowed.

Inputs:
- Task name: {{ task_name }}
- spec.md: {{ spec_md }}
- openapi.yaml: {{ openapi_yaml }}
- Candidate tests: {{ candidates_json }}

Return only JSON:
{
  "task": "{{ task_name }}",
  "decisions": [
    {
      "candidate_id": "string",
      "verdict": "KEEP or FILTER",
      "category": "SUPPORTED | OVERINTERPRETATION | SPEC_GAP | UNSUPPORTED_ASSERTION | TEST_BUG",
      "confidence": 0.0,
      "reason": "short concrete reason",
      "unsupported_assertions": ["only for FILTER"],
      "spec_support": ["visible spec/openapi support or relevant missing support"]
    }
  ]
}
\end{lstlisting}

\section{Benchmark Details}
\label{app:benchmark-details}
\label{app:task-list}

Tables~\ref{tab:task-list}--\ref{tab:task-list-final} list the 56 tasks used in the final \bench evaluation set. Each row gives the rewritten task name, coarse domain, and intended backend service boundary exposed through \texttt{spec.md} and \texttt{openapi.yaml}. Source applications were selected from publicly available open-source repositories and rewritten into benchmark tasks. We intend to release the benchmark artifacts under an appropriate research-friendly license while respecting upstream licenses.

\begin{table*}[t]
\centering
\scriptsize
\setlength{\tabcolsep}{3pt}
\renewcommand{\arraystretch}{1.03}
\begin{tabular}{@{}p{0.15\textwidth}p{0.20\textwidth}p{0.58\textwidth}@{}}
\toprule
Task & Domain(s) & Description \\
\midrule
Actual Budget & Personal finance, budgeting & Envelope budgeting service for accounts, transactions, monthly budgets, payees, scheduled transactions, and user-scoped financial data. \\
BookStack & Knowledge base, content management & Hierarchical wiki service with shelves, books, chapters, pages, page revisions, attachments, search, and permission controls. \\
Cal.com & Scheduling, calendar integration & Scheduling platform where users define availability, event types, teams, and bookings that interact with a mock calendar service. \\
Discourse & Community forum, moderation & Forum backend with categories, topics, posts, reactions, private messages, badges, and trust-level permission rules. \\
Django CRM & CRM, sales pipeline & Customer relationship management API for contacts, companies, leads, opportunities, activities, and pipeline stages. \\
Documenso & Document signing, workflow & Document signing service for documents, recipients, signature fields, sending, signing events, templates, and audit trails. \\
Dub & Link management, analytics & Short-link platform with workspaces, custom domains, tags, folders, click analytics, and role-scoped team collaboration. \\
EZBookkeeping & Personal finance, expense tracking & Expense tracking service for accounts, categories, tags, transactions, monthly summaries, budgets, transfers, and recurring entries. \\
FitTrackee & Fitness tracking, activity logging & Activity tracker for workouts, personal records, equipment, comments, GPX-like route points, and user-scoped statistics. \\
Flagsmith & Feature flags, remote config & Feature flag service for organizations, projects, environments, segments, identities, traits, and percentage rollout rules. \\
Focalboard & Project management, kanban & Kanban-style project management API for boards, cards, views, custom properties, comments, and workspace collaboration. \\
Forem & Developer community, publishing & Developer community platform for articles, comments, follows, tags, reactions, listings, and organization membership. \\
Formbricks & Surveys, audience management & Survey platform for question definitions, survey publishing, contacts, responses, segments, teams, and role-based access. \\
Ghostfolio & Portfolio tracking, wealth management & Wealth management backend for accounts, portfolio activities, holdings, benchmarks, dividends, cash flows, and performance summaries. \\
\bottomrule
\end{tabular}
\caption{\bench task list. Each row gives the task name, coarse application domain, and a one-sentence description of the required backend service.}
\label{tab:task-list}
\end{table*}

\begin{table*}[t]
\centering
\scriptsize
\setlength{\tabcolsep}{3pt}
\renewcommand{\arraystretch}{1.03}
\begin{tabular}{@{}p{0.15\textwidth}p{0.20\textwidth}p{0.58\textwidth}@{}}
\toprule
Task & Domain(s) & Description \\
\midrule
Gotify & Push notifications, admin management & Self-hosted push notification server with applications, clients, messages, user auto-creation, admin-only user management, and priority filtering. \\
GrowthBook & Feature flags, experiments & Experimentation platform for feature flags, environments, attributes, A/B experiments, metrics, snapshots, and analysis results. \\
Healthchecks & Monitoring, cron checks & Background job monitoring service where checks receive pings, track status, schedule grace periods, and trigger notification channels. \\
HedgeDoc & Markdown collaboration, access control & Collaborative markdown editor backend for notes, permissions, revisions, groups, comments, and share links. \\
Homebox & Home inventory, maintenance & Home inventory service for locations, labels, items, maintenance logs, attachments, groups, and user-scoped household data. \\
Hoppscotch & API tooling, collaboration & API development workspace for request collections, environments, teams, variables, history, and shared API testing artifacts. \\
InvenTree & Inventory, manufacturing & Inventory management system for parts, stock, suppliers, purchase orders, bills of materials, build orders, and stock movements. \\
Invoice Ninja & Invoicing, small-business finance & Invoicing platform for clients, quotes, invoices, payments, expenses, time tracking, products, and recurring billing records. \\
KitchenOwl & Recipes, grocery planning & Household recipe and grocery service with shared shopping lists, meal plans, recipes, pantry items, and expense tracking. \\
Lago & Billing, subscriptions & Usage-based billing API for customers, plans, subscriptions, meters, usage events, invoices, payments, and coupons. \\
Langfuse & LLM observability, evaluation & Observability backend for LLM traces, generations, prompts, datasets, evaluation scores, projects, and API keys. \\
Leantime & Project management, OKRs & Goal-focused project management service with goals, milestones, tasks, sprints, timesheets, ideas, and project-level permissions. \\
Linkwarden & Bookmark management, collaboration & Collaborative bookmark manager for collections, links, tags, archive metadata, sharing, and user or team access controls. \\
Logto & Identity management, authorization & Identity platform management API for users, applications, API resources, permissions, roles, organizations, and sign-in settings. \\
\bottomrule
\end{tabular}
\caption{\bench task list, tasks 15--28.}
\label{tab:task-list-b}
\end{table*}

\begin{table*}[t]
\centering
\scriptsize
\setlength{\tabcolsep}{3pt}
\renewcommand{\arraystretch}{1.03}
\begin{tabular}{@{}p{0.15\textwidth}p{0.20\textwidth}p{0.58\textwidth}@{}}
\toprule
Task & Domain(s) & Description \\
\midrule
Maybe Finance & Personal finance, analytics & Personal finance dashboard for accounts, transactions, categories, budgets, goals, net-worth views, and spending insights. \\
RecipeVault & Recipes, meal planning & Self-hosted recipe manager inspired by Mealie with recipes, categories, tags, meal plans, shopping lists, ratings, and households. \\
Medusa & E-commerce, order management & Headless commerce backend for products, customers, carts, orders, payments, discounts, shipping, and regional configuration. \\
Memos & Notes, social publishing & Lightweight note-taking service for markdown memos, tags, visibility, resources, reactions, comments, and user timelines. \\
Miniflux & RSS reader, content aggregation & RSS reader backend for feeds, categories, entries, bookmarks, read state, subscriptions, and feed refresh behavior. \\
NetBox & Infrastructure management, IPAM & Network and datacenter management API for sites, racks, devices, interfaces, cables, prefixes, IP addresses, and tenants. \\
NocoDB & No-code database, spreadsheet & No-code database API for bases, tables, typed columns, records, views, filters, comments, webhooks, and audit logs. \\
Novu & Notifications, workflow orchestration & Notification infrastructure backend for workflows, subscribers, topics, templates, and mock multi-channel delivery behavior. \\
Ntfy & Push notifications, pub/sub & Topic-based notification service for publishing messages, subscriptions, priorities, attachments, tokens, and access policies. \\
Outline & Knowledge base, document collaboration & Collaborative wiki backend for collections, nested documents, comments, search, sharing, groups, and access permissions. \\
Paperless-ngx & Document management, metadata & Document management service for correspondents, document types, tags, custom fields, saved views, tasks, and bulk edits. \\
Plane & Project management, issue tracking & Project management backend for workspaces, projects, workflow states, issues, cycles, modules, views, pages, and comments. \\
Plausible & Web analytics, event tracking & Privacy-first analytics API for sites, pageviews, custom events, goals, aggregate metrics, and API-key-scoped reports. \\
Saleor & E-commerce, inventory and checkout & Multi-channel commerce backend for products, attributes, warehouses, stock, checkouts, payments, orders, and fulfillment flows. \\
\bottomrule
\end{tabular}
\caption{\bench task list, tasks 29--42.}
\label{tab:task-list-c}
\end{table*}

\begin{table*}[t]
\centering
\scriptsize
\setlength{\tabcolsep}{3pt}
\renewcommand{\arraystretch}{1.03}
\begin{tabular}{@{}p{0.15\textwidth}p{0.20\textwidth}p{0.58\textwidth}@{}}
\toprule
Task & Domain(s) & Description \\
\midrule
Shlink & URL shortener, analytics & Self-hosted URL shortener with short codes, redirects, tags, visit tracking, domains, QR codes, and rule-based access. \\
Solidtime & Time tracking, reporting & Time-tracking service for organizations, projects, tasks, time entries, approvals, clients, and aggregate reports. \\
Spree & E-commerce, multi-store checkout & Multi-store commerce platform with products, taxonomies, carts, orders, promotions, shipping, stock, and checkout state transitions. \\
Strapi & Headless CMS, content modeling & Headless CMS API for dynamic content types, entries, relations, components, media, localization, and role-based permissions. \\
Tandoor Recipes & Recipes, meal planning & Recipe management backend for recipes, ingredients, meal plans, shopping lists, nutrition metadata, and cookbook organization. \\
TaskCafe & Project management, kanban & Kanban project management service for projects, task groups, tasks, labels, checklists, comments, and member assignments. \\
Traduora & Translation management, localization & Localization platform for projects, locales, translation keys, translated values, collaborators, exports, and progress tracking. \\
TubeArchivist & Media archiving, video metadata & Media archive API for videos, channels, playlists, download queues, comments, subtitles, sponsor segments, and metadata search. \\
Twenty & CRM, workspace data model & Workspace CRM backend for people, companies, opportunities, pipelines, custom objects, activities, notes, and team roles. \\
Umami & Web analytics, tracking & Privacy-focused analytics service for websites, public event collection, sessions, metrics, reports, and team-scoped dashboards. \\
Vaultwarden & Password manager, encrypted vault & Bitwarden-compatible vault service for ciphers, folders, collections, organizations, sends, attachments, and token-based access. \\
Vikunja & Task management, collaboration & Self-hosted task management API for projects, nested tasks, labels, due dates, reminders, sharing, and comments. \\
Zitadel & Identity management, access control & Multi-tenant identity platform for organizations, users, sessions, projects, roles, MFA, password policies, and OAuth-style resources. \\
Zulip & Team chat, topic-based messaging & Topic-based team chat service with streams, topics, messages, reactions, users, subscriptions, private messages, and moderation rules. \\
\bottomrule
\end{tabular}
\caption{\bench task list, tasks 43--56.}
\label{tab:task-list-final}
\end{table*}